\documentclass[manuscript]{aastex}

\slugcomment{}

\shorttitle{MHD waves in the solar tachocline} \shortauthors{Zaqarashvili et
al.}

\begin{document}


\title{Global shallow water magnetohydrodynamic waves in the solar tachocline}


\author{T.V. Zaqarashvili}
\affil{Abastumani Astrophysical Observatory at I. Chavchavadze State
University, Kazbegi Ave. 2a, Tbilisi 0160, Georgia}
\email{temury@genao.org}

\author{R. Oliver and J.L. Ballester}
\affil{Departament de F\'{\i}sica, Universitat de les Illes Balears,
E-07122 Palma de Mallorca, Spain}
\email{[ramon.oliver;dfsjlb0]@uib.es}

%\altaffiltext{1}{Also at Abastumani Astrophysical Observatory, Kazbegi Ave. 2a, Tbilisi 0160, Georgia}



\begin{abstract}

We derive analytical solutions and dispersion relations of global
magnetic Poincar\'e (magneto-gravity) and magnetic Rossby waves in
the approximation of shallow water magnetohydrodynamics. The
solutions are obtained in a rotating spherical coordinate system for
strongly and weakly stable stratification separately in the presence
of toroidal magnetic field. In both cases magnetic Rossby waves
split into fast and slow magnetic Rossby modes. In the case of
strongly stable stratification (valid in the radiative part of the
tachocline) all waves are slightly affected by the layer thickness
and the toroidal magnetic field, while in the case of weakly stable
stratification (valid in the upper overshoot layer of the
tachocline) magnetic Poincar\'e and magnetic Rossby waves are found
to be concentrated near the solar equator, leading to equatorially
trapped waves. The frequencies of all waves are smaller in the upper
weakly stable stratification region than in the lower strongly
stable stratification one.

\end{abstract}


\section{Introduction}

The tachocline, which is believed to exist near the base of the solar
convection zone \citep{spi92}, has become a subject of rapidly growing
scientific interest in the last years \citep{gou07}. The existence of
the tachocline is important for two main reasons. First, it may
prevent the spread of the solar angular momentum from the convection
zone to the interior \citep{spi92,gou98,gar07}, thus retaining the
inferred solid body rotation in the radiative zone; second, the
processes of solar magnetic field generation and global
magnetohydrodynamic (MHD) instabilities, which are crucial for
solar activity, may take place there \citep{dik05,gil07}. MHD waves
and oscillations in the tachocline may play a significant role in both
cases: they may redistribute the angular momentum in the horizontal
direction and some of them may become unstable due to the
differential rotation, leading to magnetic flux emergence at the
solar surface. Therefore, to study the complete spectrum of possible
wave modes in this system is of vital importance. The tachocline is
very thin compared to the solar radius, therefore the ordinary shallow
water approximation modified by the presence of a horizontal
large-scale magnetic field can be easily applied \citep{gil00}.

The spectrum of various shorter scale shallow water MHD waves has
been recently studied in Cartesian coordinates by \citet{sch01}.
However, global wave modes, those with a wavelength comparable to
the solar radius, must be considered in spherical coordinates.
Recently, \citet{zaq07} have studied the spherical shallow water MHD waves
in the simplest case, in which the high frequency branch i.e. magnetic
Poincar\'e waves (or magneto-gravity waves) and the influence of the
tachocline thickness on the wave dynamics have been ignored (that is,
when the surface gravity speed is much higher than the surface
rotation speed). This approximation has enabled us to study the dynamics
of global magnetic Rossby waves in the lower, strongly stable part of
the tachocline, but it fails in the upper, weakly stable overshoot
region, where a negative buoyancy due to the subadiabatic
stratification strongly reduces the gravity speed \citep{gil00}.

In this letter, we present the analytical spectrum of global linear
shallow water MHD waves for both parts of the tachocline in the
absence of differential rotation.

\section{Global shallow water MHD waves}

We use the linearized shallow water MHD equations in the rotating
spherical coordinate system ($r,\theta, \phi$), where $r$ is the
radial coordinate, $\theta$ is the co-latitude, and $\phi$ is the
longitude \citep{zaq07}. We also assume the unperturbed toroidal
magnetic field to be $B_{\phi}=B(\theta)\sin{\theta}$, where
$B(\theta)$ may have an arbitrary profile with latitude, which can
be chosen later.
%This is the
%simplest profile since a more complex structure of the magnetic
%field leads to the coupling of different harmonics and consequently
%an analytical dispersion relation cannot be obtained.
Then a normal mode analysis of the form $\exp(-i\omega t + is\phi)$
gives the equation \citep{zaq07}
\begin{eqnarray}
(\lambda D +s \mu)\left \{{1\over {s^2 -\epsilon
\lambda^2(1-\mu^2)}}\left [\lambda D -s \mu- {{\alpha^2}\over
{\lambda^2}}s^2\lambda(D +2\mu)\right ]\right \}{\hat u}_{\theta}
-(\lambda^2 -\mu^2){\hat u}_{\theta} \nonumber
\\ +s^2\alpha^2{\hat u}_{\theta} + 2\alpha^2\mu D{\hat u}_{\theta}
+\mu s {\alpha^2\over \lambda}(D+2\mu){\hat u}_{\theta}+\mu
D(\alpha^2) {\hat u}_{\theta}=0,
\end{eqnarray}
with
\begin{eqnarray*}
\lambda={{\omega}\over {2\Omega_0}},\,
\epsilon={{4\Omega^2_0R^2_0}\over {g H_0}},\,\alpha^2={{v^2_A}\over
{4\Omega^2_0R^2_0}},\,\,v_A={{B(\mu)}\over {\sqrt {4\pi
\rho}}},\nonumber
\\  \mu=\cos \theta,\,\,D=(1-\mu^2){{\partial
}\over {\partial \mu}}=-\sin \theta{{\partial }\over {\partial
\theta}}, \,\, {\hat u}_{\theta}=\sin {\theta} u_{\theta},
\end{eqnarray*}
where $u_{\theta}$ is the poloidal velocity component, $g$ is the
reduced gravity in the tachocline, $\rho$ is the medium density,
$\Omega_0$ is the angular velocity, $R_0$ is the distance from the
solar center to the tachocline, and $H_0$ is the tachocline
thickness. This is the general equation for the linear dynamics of
the shallow water MHD system, which contains various kinds of waves
(magnetic Poincar\'e and magnetic Rossby waves). When $\epsilon=0$
and $B=const$, this equation transforms into the associated Legendre
equation and governs the dynamics of fast and slow magnetic Rossby
waves \citep{zaq07}. This approximation neglects magnetic Poincar\'e
waves and the influence of the tachocline thickness on the dynamics
of magnetic Rossby waves. Therefore, equation (1) must be solved
with $\epsilon \not=0$ in order to give the complete spectrum of
waves. The subadiabatic stratification in the tachocline provides
negative buoyancy, which leads to a reduced gravity and consequently
an increased value of $\epsilon$. In the lower strongly stable part
of the tachocline $\epsilon$ still is much smaller than unity, but
in upper overshoot layer $\epsilon$ becomes much greater than unity.
Therefore, the solution of equation (1) for the two extreme cases
$\epsilon \ll 1$ and $\epsilon \gg 1$ covers both parts of the
tachocline. Here we use a magnetic field profile typical of the
Sun, namely
\begin{equation}B({\mu})=B_0 \mu,
\end{equation}
which has opposite signs in the northern and southern hemispheres
\citep{gil97}. The magnetic field profile (2) leads to
$\alpha=\alpha_0 \mu$, where $\alpha_0=B_0/(2\Omega_0 R_0 \sqrt{4\pi
\rho})$. In the next subsections we will give analytical solutions
and dispersion relations of global shallow water MHD waves for each
case separately.

\subsection{$\epsilon \ll 1$ (strongly stable
stratification; valid for the radiative part of tachocline)}

Reduced gravity in the strongly stable radiative part of the
tachocline can be estimated as 500--1.5 $\cdot$ 10$^4$
cm$\cdot$s$^{-2}$ \citep{sch01}. Then, for the layer thickness
$H_0$=10$^9$ cm we get $\epsilon=$ 4.5 $\cdot$ 10$^{-3}$ -- 0.13
(where $\Omega_0$=2.6 $\cdot$ 10$^{-6}$ s$^{-1}$, $R_0$=5$\cdot$
10$^{10}$ cm have been used). Therefore, $\epsilon \ll 1$ is a good
approximation in the radiative part of the tachocline.

In this approximation and using the weak magnetic field limit, i.e.
$\alpha_0^2 \ll 1$, equation (1) leads to the spheroidal wave
equation (second order terms with $\epsilon$ and $\alpha_0^2$ are
neglected), whose typical solutions are the spheroidal wave
functions $S_{sn}$ \citep{abr64}, where $n$ plays the role of the
poloidal wavenumber. The approximate solution can be written as
\begin{equation}
{\hat u}_{\theta} \propto u_0 S_{sn}({\epsilon_1}, \mu)\sqrt {{s^2
-\epsilon \lambda^2(1-\mu^2)}\over {\lambda^2 - s^2\alpha^2_0
\mu^2}},
\end{equation}
where $u_0$ is the amplitude and $$\epsilon_1=\sqrt {\epsilon \left
(1 + 3{\lambda^2\over s^2} -2 {\lambda\over s}\right ) +
{\alpha^2_0s^2\over \lambda^2}\left (7 + {s\over {2\lambda}} \right
)}.$$ Note that $n$ is integer for the solutions, which vanish at
the poles ($\mu=\pm 1$). The dispersion relation is found to be
\begin{equation}\label{dr_small_eps}
\epsilon (s^2 +1)\left ({{\lambda}\over s}\right )^4 -n(n+1)\left
({{\lambda}\over s}\right )^2 - \left ({{\lambda}\over s}\right ) +
\alpha^2_0 =0.
\end{equation}

The dispersion relation can be split into high (magnetic Poincar\'e)
and low (magnetic Rossby) frequency branches. For the magnetic
Poincar\'e waves we have
\begin{equation}\label{approx_mp_small_eps}
{{\lambda_{mP}}\over s} \approx \pm \sqrt { {{n(n+1)}\over {\epsilon
(s^2 +1)}} }.
\end{equation}
This is the dispersion relation of ordinary surface gravity waves
and we can recover it for the high $s$ harmonics \citep{lon65}.

For the magnetic Rossby waves we have approximately
\begin{equation}
n(n+1)\left ({{\lambda}\over s}\right )^2 + {{\lambda}\over s} -
\alpha^2_0=0.
\end{equation}

\noindent It is clear from equation (6) that the magnetic field causes the
splitting of ordinary Rossby waves into fast and slow modes
\citep{zaq07}. For $\alpha_0 \ll 1$ the dispersion relations for the
fast and slow magnetic Rossby modes can be approximated as
\begin{equation}\label{approx_fs_small_eps}
{{\lambda_f}\over s}\approx -{1\over {n(n+1)}} - \alpha^2_0,\,\,\,
{{\lambda_s}\over s}\approx \alpha^2_0.
\end{equation}
We see that the fast magnetic Rossby mode has a dispersion
relation similar to that of the ordinary Rossby waves but slightly modified
by the magnetic field. Slow magnetic Rossby waves, however, are new
wave modes as their dispersion relation is different from Rossby and
Alfv\'en wave dispersion relations (the dispersion relation of
Alfv\'en waves should be $\lambda_A={\pm}s\alpha_0$). There is a
significant difference between the frequencies of fast and slow
magnetic Rossby modes for sufficiently small $\alpha_0$.

The dispersion diagram for the $s=1$ harmonics of shallow water MHD
waves according to Eq. (\ref{dr_small_eps}) is shown in Fig 1. Here $\epsilon = 0.01$
and $\alpha_0=0.05$. The frequencies of magnetic Poincar\'e waves
(solid lines) are much higher than the rotational frequency,
$\Omega_0$, and they increase with increasing $n$. For example, the
frequency of the $n=3$ harmonics is $\sim 50 \, \Omega_0$, which
corresponds to oscillations with period $\sim$ 13 hours.  On the
other hand, the frequencies of magnetic Rossby waves are in general much lower
than the rotational frequency. The absolute value of the frequency of fast magnetic Rossby
waves (dotted line in the zoom) decreases with increasing $n$, such
as happens with the HD Rossby wave.

Fig. 2 shows the dependence of the $s=1, n=2$ and $s=1, n=3$
harmonics on $\epsilon$ and $\alpha_0$ (i.e. the magnetic field
strength). The frequency (in absolute value) of fast (dotted) and slow (dashed) magnetic
Rossby modes significantly increases when the magnetic field is
increased.  We also see that the frequency difference between the $n=2,3$ harmonics of fast magnetic
Rossby waves is almost independent of $\alpha_0$. On the contrary, the $n=2,3$ harmonics of slow magnetic
Rossby waves diverge when the magnetic field is increased. It turns
out that the frequencies of magnetic Rossby waves have almost no
dependence on $\epsilon$, therefore they are not shown in the top panel. On the
other hand, the frequency of magnetic Poincar\'e waves depends on
$\epsilon$ (lower panel) and significantly decreases with increasing
$\epsilon$. However, these frequencies have almost no dependence on
$\alpha_0$, therefore magnetic Poincar\'e waves are only slightly
affected by the magnetic field.

\subsection{$\epsilon \gg 1$ (weakly
stable stratification; valid in the upper overshoot part of
tachocline)}

In the weakly stable overshoot region (or say upper tachocline)
reduced gravity is much smaller, being 0.05--5 cm$\cdot$s$^{-2}$
\citep{sch01}. Then, for the thickness $H_0$=5 $\cdot$ 10$^8$ cm we
get $\epsilon=$ 27 -- 2.7 $\cdot$ 10$^{3}$. Therefore, $\epsilon \gg
1$ is a good approximation in the overshoot region.

In this case, we follow the calculation of \citet{lon68} (this paper
considers spherical hydrodynamic shallow water waves in the Earth
context), and introduce a new function $\eta = \epsilon^{1/4} \mu$
and also consider that $\lambda=\epsilon^{-1/4}L$ and
$\alpha_0=\epsilon^{-1/4} L$, where $L$ is of order unity. Then
after some algebra and keeping only large terms equation (1) leads
to the Weber (parabolic cylinder) equation, whose solutions finite
at $\eta \rightarrow \pm \infty$ can be expressed in terms of
Hermite polynomials $H_{\nu}(\eta)$ of order $\nu$, where $\nu$
satisfies
\begin{equation}
{{{\epsilon} \lambda^2 - {{s}\over {\lambda}} + {{\alpha_0^2
s^2}\over {\lambda^2}}}}=(2 \nu + 1)\sqrt {\epsilon},
\,\,\hspace{1cm} (\nu =0,1,2,\ldots)
\end{equation}
and the solution is proportional to
\begin{equation}
{\hat u}_{\theta} \propto e^{-{1\over 2}\eta^2}H_{\nu}(\eta).
\end{equation}
Therefore the function ${\hat u}_{\theta}$ is exponentially small
beyond the turning points $\cos \theta= \epsilon^{-1/4} \sqrt {2 \nu
+ 1}$. Due to the large $\epsilon$, the co-latitude $\theta$
satisfying this condition is close to $90^{\rm o}$, therefore the
solution is confined to the neighborhood of the equator.

Equation (8) yields the wave dispersion relation as
\begin{equation}
\lambda^4 - {{2\nu +1}\over {\sqrt {\epsilon}}}\lambda^2 -{s\over
{\epsilon}} \lambda + {{\alpha_0^2 s^2}\over \epsilon}=0.
\end{equation}
In the non-magnetic case this equation transforms into equation
(8.11) of \citet{lon68}. This is a fourth order equation and it
describes magnetic Poincar\'e and magnetic Rossby waves. For large
values of $\epsilon$, the solutions can be written separately for
both types of waves.

The solutions for Poincar\'e waves can be written approximately as
\begin{eqnarray}\label{approx_mp_large_eps}
\lambda_{mP} \approx \pm {{(2\nu +1)^{1/2}}\over {\epsilon^{1/4}}} +
{s\over {(4\nu +2)\epsilon^{1/2}}}.
\end{eqnarray}
It turns out that the magnetic field has almost no influence on the
dynamics of Poincar\'e waves and the dispersion relation is the same
as for the nonmagnetic case \citep{lon68}.

The dispersion relation for the magnetic Rossby waves can be written
as
\begin{equation}
 (2\nu +1)\sqrt {\epsilon}\lambda^2 +s \lambda - \alpha_0^2 s^2=0.
\end{equation}
The magnetic field causes the splitting of ordinary Rossby waves
into fast and slow magnetic Rossby waves as in the $\epsilon \ll 1$
case. If $\sqrt {\epsilon}\alpha_0^2 \ll 1$, then we can write
\begin{eqnarray}\label{approx_fs_large_eps}
{\lambda_f \over s} \approx  - {1 \over {(2\nu +1)\epsilon^{1/2}}}
-\alpha_0^2, \,\,\, {\lambda_s \over s} \approx \alpha_0^2.
\end{eqnarray}
Fast magnetic Rossby waves are similar to hydrodynamic Rossby
waves modified by the magnetic field. Slow magnetic Rossby waves
have a similar dispersion relation as in the $\epsilon \ll 1$ case.

The dispersion diagram for the $s=1$ harmonics of magnetic
Poincar\'e and magnetic Rossby waves according to Eq. (10) is shown
in Fig 3. Here $\epsilon = 2700$ and $\alpha_0=0.05$. The behaviour
of fast magnetic Rossby waves (dotted line) is similar to the small
$\epsilon$ case: the absolute value of their frequency significantly decreases with
increasing $\nu$ (which now plays the role of the poloidal wave
number). However, we immediately note that the frequency now is much
smaller than that of the $\epsilon\ll 1$ case. This is also true for
magnetic Poincar\'e waves (solid lines): their frequency is
significantly small compared to the case of small $\epsilon$, now
being even smaller than the rotational frequency. The frequency of
slow magnetic Rossby waves only slightly depends on $\nu$, but now
it decreases with increasing $\nu$, contrary to $\epsilon\ll 1$
case.

Fig. 4 shows the dependence of the $s=1, \nu=2$ and $s=1, \nu=3$
harmonics on $\epsilon$ and $\alpha_0$. The frequency of magnetic
Rossby mode harmonics significantly depends on $\alpha_0$; the
frequencies of both fast and slow magnetic Rossby modes increase
with increasing $\alpha_0$. On the other hand, it turns out that the
frequencies of magnetic Poincar\'e waves have almost no dependence
on $\alpha_0$. However, they are significantly reduced with
increasing $\epsilon$, becoming smaller than the rotational
frequency for $\epsilon > 1000$. Fast magnetic Rossby waves also
depend on $\epsilon$, their frequency significantly decreasing when
this parameter is increased. Slow magnetic Rossby waves have almost no dependence on $\epsilon$
as it is suggested from Eq. (\ref{approx_fs_large_eps}). Thus increasing $\epsilon$, which
is equivalent to reducing $g$, generally leads to a decrease of the
wave frequencies.

The numerical solution of the dispersion relation (10) shows that
the frequency of the magnetic Rossby wave harmonic with $\nu=0$ becomes
complex for sufficiently large $\alpha_0$, i.e. large magnetic
field. This may point to some kind of instabilities similar to the
polar kink instability found by Cally (2003). However, the value of
$\alpha_0$ for which the frequency becomes complex is significantly
outside the range for which Eq. (10) is valid (note that $\alpha_0$
should be proportional to $\epsilon^{-1/4}$). Therefore, the complex
solution can be spurious and may not reflect a real instability.

%In order to study the slow magnetic Rossby waves, one should
%consider the very low frequency branch so that $\lambda^2 <
%\alpha^2s^2$. We again follow \citet{lon68}, but now consider
%$\lambda=\epsilon^{-1/2}L$ instead of $\lambda=\epsilon^{-1/4}L$.
%Then it turns out that the solutions are concentrated near the poles
%and are linked to the associated Laguerre polynomials as
%\begin{equation}
%{\hat u}_{\theta} \propto x^{{{1}\over 2}s}e^{{1\over
%2}x}L^{s}_{\nu}(x),
%\end{equation}
%where
%\begin{equation}
%x=\sqrt {A_2} \theta^2,\,\,\, L^{s}_{\nu}(x)=\sum^{{\nu}}_{k=0}
%{{(s+{\nu})!}\over {(s+k)!({\nu}-k)!k!}}(-x)^k,\,\,\, A_2=
%-{{{\epsilon} \lambda^2}\over {\lambda^2-\alpha^2s^2}} +{1\over
%4}{{\epsilon^2 \alpha^4 \lambda^2 s^2}\over
%{(\lambda^2-\alpha^2s^2)^2}},
%\end{equation}
%so that $s>-1$.
%
%The dispersion relation for slow magnetic Rossby waves is
%\begin{equation} {\lambda\over s}= - {1\over
%\epsilon}\left (1 + 3{{\epsilon \alpha^2}}  \right ) +
%2({{2{\nu}+s+1}}){\alpha\over {\sqrt {\epsilon}}}\sqrt {1 +
%{{\epsilon \alpha^2}\over 4}}.
%\end{equation}

\section{Conclusions}

We have derived analytical solutions and dispersion relations of
global shallow water MHD waves for the solar tachocline in a
rotating spherical coordinate system. The solutions and dispersion
relations are obtained for weakly and strongly stable
stratifications separately. The weakly stable stratification is
valid in the upper overshoot part of tachocline, while the strongly
stable stratification is valid for the radiative part of tachocline.
The solutions include the Poincar\'e and Rossby waves modified by
the presence of the magnetic field.

Here we use a realistic latitudinal profile of the toroidal magnetic
field $B_{\phi}\sim \cos{\theta} \sin{\theta}$ that changes
sign at the equator. Similar to the $\epsilon=0$ case \citep{zaq07}, the
magnetic field leads to the splitting of hydrodynamic Rossby waves
into fast and slow magnetic Rossby modes.

In the case of strongly stable stratification, the dispersion
relation of magnetic Poincar\'e waves is similar to that of the
hydrodynamic case \citep{lon65}. The dispersion relation of magnetic
Rossby waves \citep{zaq07} is slightly modified by the thickness of
the layer. In the case of weakly stable stratification, both
magnetic Poincar\'e and magnetic Rossby waves are concentrated near
the equator, thus leading to equatorially trapped waves.

Another remarkable feature is that the frequencies of all waves are
significantly reduced in the large $\epsilon$ case compared to the
$\epsilon\ll 1$ one. This means that the upper overshoot part of the
tachocline supports oscillations with smaller frequency than the
lower radiative part of the tachocline.

The solutions have been obtained without taking into account the
differential rotation which is present in the tachocline. The future
inclusion of differential rotation probably will lead to the
instability of some modes, which was found by simulations in the
inertial frame \citep[][and references therein]{gil07}.

%The magnetic Rossby waves may have interesting consequences in
%intermediate periodicity found in the activity of the Sun and other
%stars.

\acknowledgments

This work has been supported by MICINN grant AYA2006-07637 and
FEDER funds, by the Conselleria de Economia, Hisenda i Innovaci\'o under grant
PCTIB-2005GC3-03, and by the
Georgian National Science Foundation grant GNSF/ST06/4-098.

\clearpage

\begin{figure}
\includegraphics[angle=0,scale=0.8]{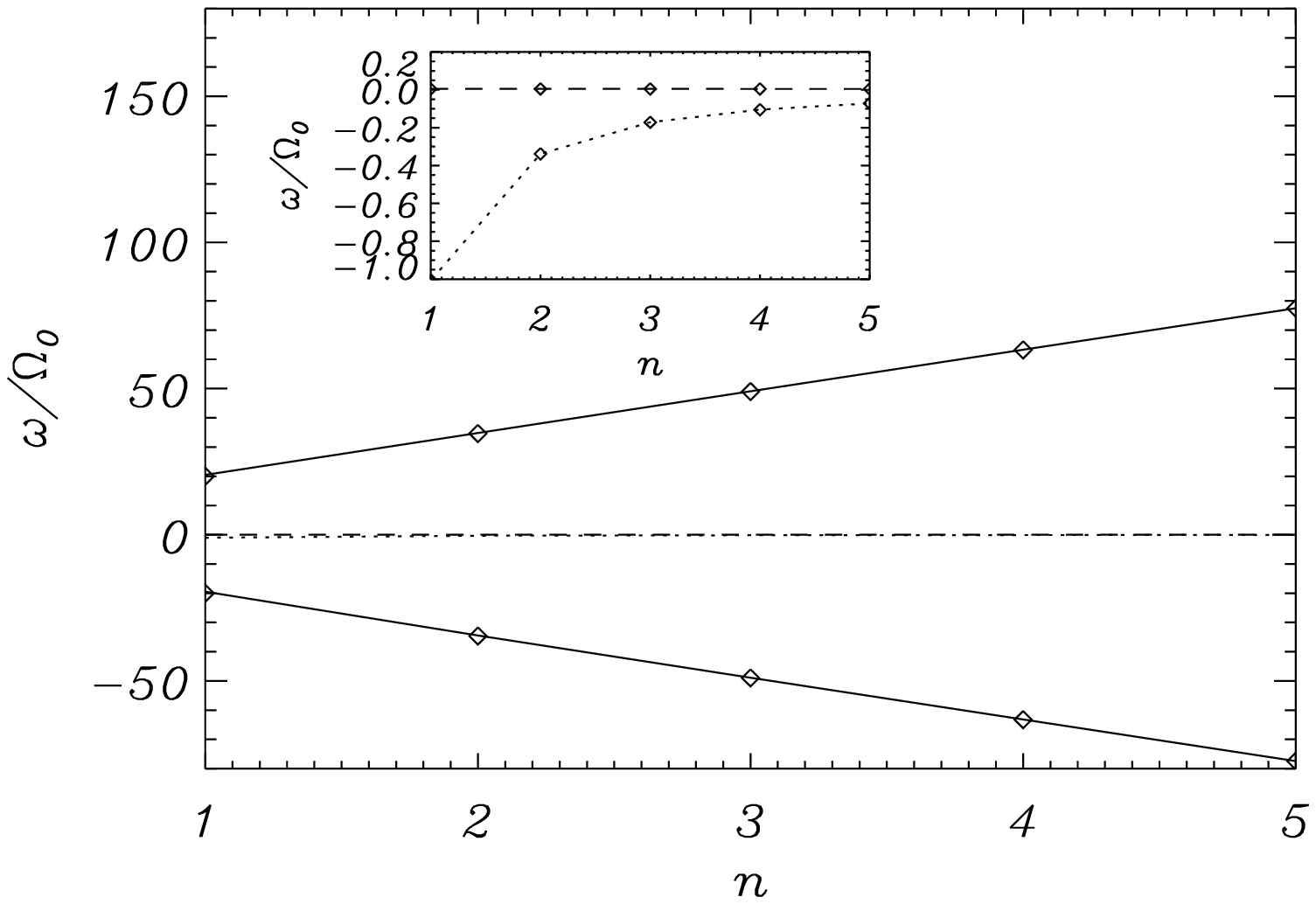} \caption{Dispersion diagram of spherical shallow water waves in the
presence of a toroidal magnetic field for small $\epsilon$. The two
extreme upper and lower solutions (solid lines) correspond to
magnetic Poincar\'e waves, whereas the two low frequency modes shown
also in the zoom are fast (dotted) and slow (dashed) magnetic Rossby
waves. Symbols denote solutions obtained with the approximate
formulas (\ref{approx_mp_small_eps}) and (\ref{approx_fs_small_eps}). The parameters used to obtain the dispersion
diagram are $\epsilon =0.01$, $s=1$, and $\alpha_0=0.05$. }
\end{figure}

\begin{figure}
\includegraphics[angle=0,scale=0.8]{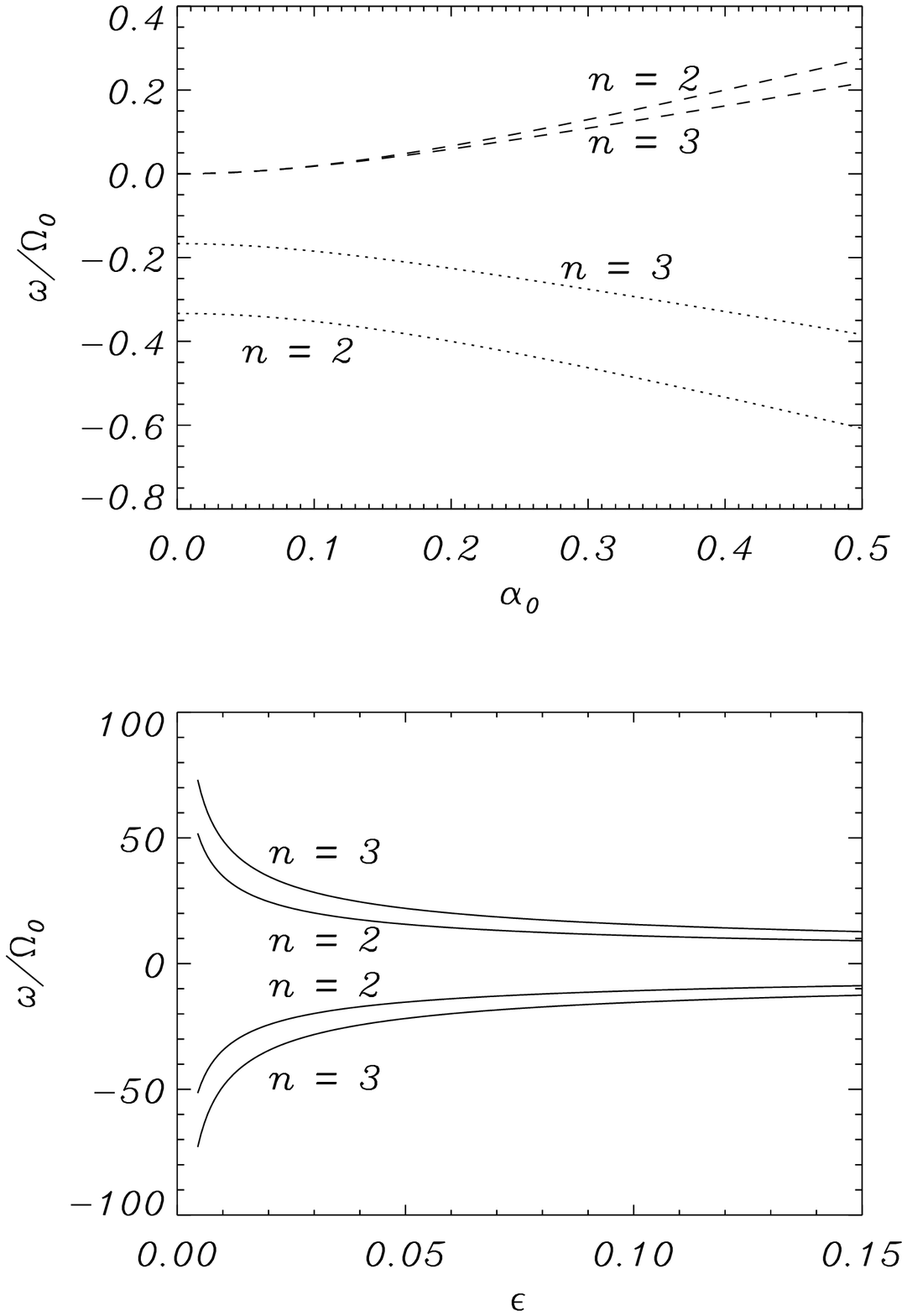} \caption{Upper panel:
frequencies of the $s=1, n=2$ and $s=1, n=3$ harmonics of fast
(dotted) and slow (dashed) magnetic Rossby modes vs. $\alpha_0$ for
$\epsilon=0.0045$. Magnetic Poincar\'e waves do not depend
significantly on the magnetic field, therefore they are not shown
here. Lower panel: frequencies of the $s=1, n=2$ and $s=1, n=3$
harmonics of magnetic Poincar\'e waves vs. $\epsilon$ for
$\alpha_0=0.05$. }
\end{figure}

\begin{figure}
\includegraphics[angle=0,scale=0.8]{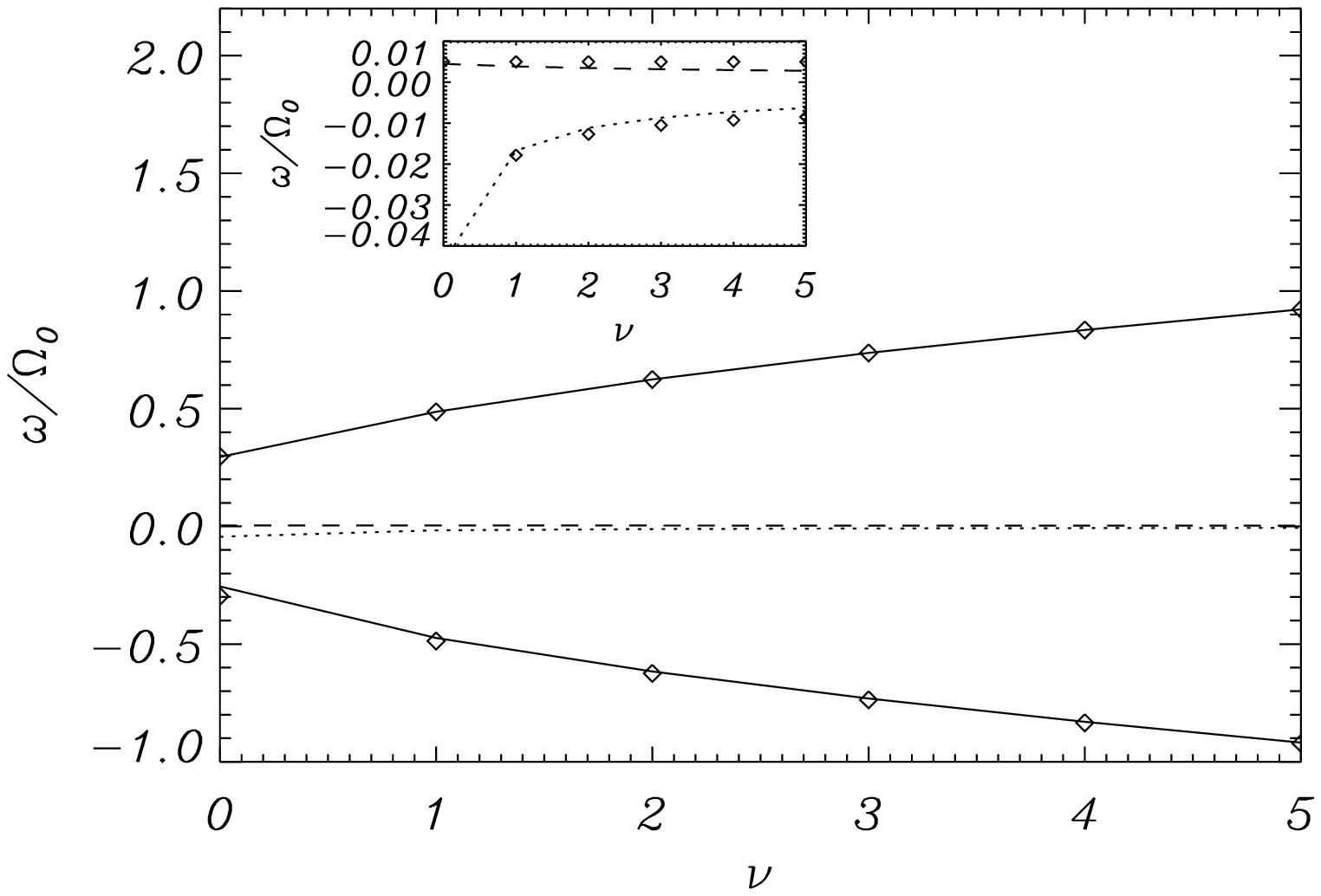} \caption{Dispersion
diagram of spherical shallow water waves in the presence of a
toroidal magnetic field for large $\epsilon$. The two extreme upper
and lower solutions (solid lines) correspond to magnetic Poincar\'e
waves, whereas the low frequency modes (also shown with zoom) are
the fast (dotted) and slow (dashed) magnetic Rossby waves. Symbols
denote solutions obtained with the approximate formulas (\ref{approx_mp_large_eps}) and (\ref{approx_fs_large_eps}). The
parameters used to obtain the dispersion diagram are $\epsilon
=2700$, $s=1$, and $\alpha_0=0.05$. }
\end{figure}

\begin{figure}
\includegraphics[angle=0,scale=0.7]{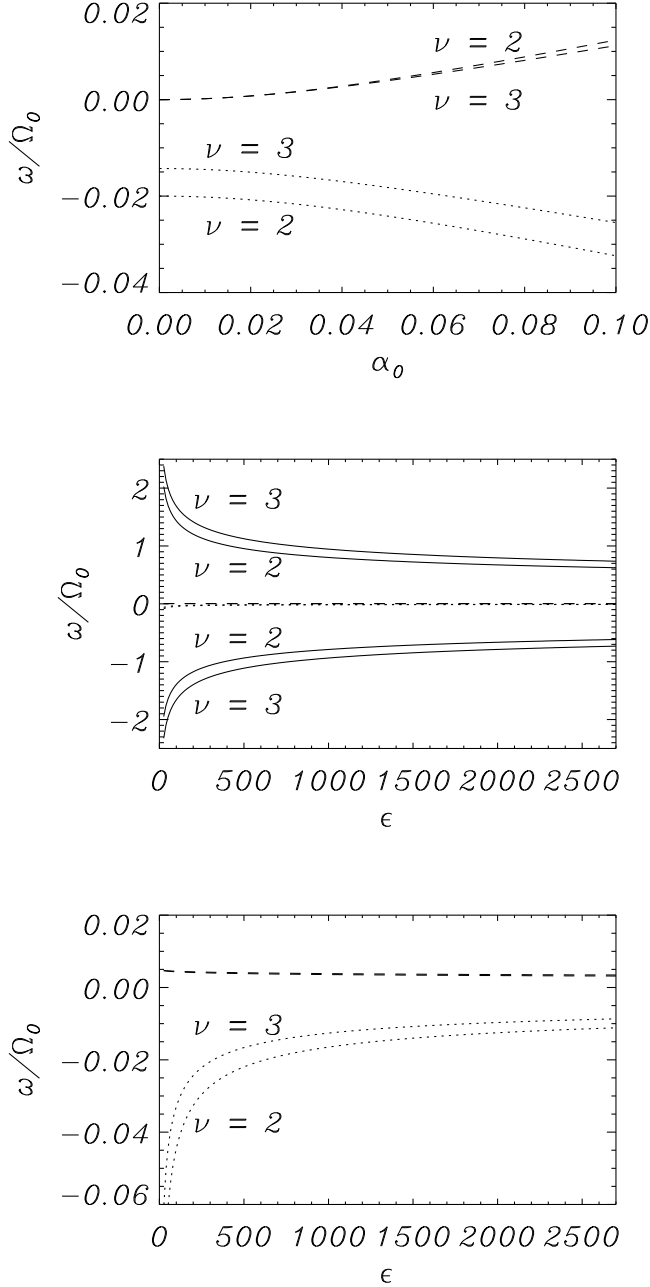} \caption{Upper panel:
frequencies of the $s=1, \nu=2$ and $s=1, \nu=3$ harmonics of fast
(dotted) and slow (dashed) magnetic Rossby modes vs. $\alpha_0$ for
$\epsilon=400$. Middle panel: frequencies of the $s=1, \nu=2$ and
$s=1, \nu=3$ harmonics of magnetic Poincar\'e and magnetic Rossby
modes vs. $\epsilon$ for $\alpha_0=0.05$. Lower panel: same as in
middle panel, but only for magnetic Rossby waves. }
\end{figure}

\end{document}